\documentclass[multphys,vecphys]{svmult}

\usepackage{makeidx}         
\usepackage{graphicx}        
\usepackage{multicol}        
\usepackage[bottom]{footmisc}

\makeindex             

\begin{document}

\title*{Abundance Anomalies in Galactic Globular Clusters - Looking 
for the Stellar Culprits}
\titlerunning{Abundance Anomalies in Galactic GCs}
\author{C.Charbonnel\inst{1,2} \& N.Prantzos\inst{3}}
\institute{Geneva Observatory, 51, chemin des Maillettes, CH-1290 Sauverny, Switzerland
\texttt{Corinne.Charbonnel@obs.unige.ch}
\and LATT, CNRS UMR 5572, 14, av.E.Belin, 31400 Toulouse, France
\and IAP, CNRS UMR 7095, 98b Bd.Arago, 75014 Paris, France
\texttt{prantzos@iap.fr}}
%
%
\maketitle

\abstract{Galactic globular cluster stars exhibit abundance patterns which 
are not shared by their field counterparts. It is clear from recent 
spectroscopic observations of GC turnoff stars that these abundance 
anomalies were already present in the gas from which the observed stars 
formed. This provides undisputed support to the so-called self-enrichment 
scenario according to which a large fraction of GC low-mass stars have formed
from material processed through hydrogen-burning at high temperatures and 
then lost by more massive and faster evolving stars (and perhaps mixed with 
some original gas). Within this framework we present a new method to 
derive the Initial Mass Function of the polluter stars.}

\section{Abundance anomalies in galactic globular clusters}
\label{sec:1}
During the last three decades, an incredible amount of data
has been collected on the chemical properties of galactic globular clusters
(hereafter GCs) thanks to high spectral resolution abundance analysis
(see \cite{Grattonetal2004} and \cite{Sneden2005} for recent reviews).
The main results can be summarized as follows : (i) Individual GCs (with the notable 
exception of Omega Cen) appear to be fairly homogeneous as far as the iron peak elements 
(Ni, Cu) are concerned;
(ii) They present very low scatter and the same trends as field stars
for the neutron-capture elements (Ba, La, Eu) and the alpha-elements (Si, Ca);
(iii) They exhibit however complex patterns and large star-to-star abundance
variations for the lighter elements from C to Al which are not shared by
their field counterparts. 
Among these anomalous patterns, the most striking ones are the so-called 
universal O-Na anticorrelation (first discovered by the Lick-Texas group) 
and the Mg-Al anticorrelation 
(see e.g. \cite{RamirezCohen2002} and \cite{Ivansetal1999}).
These patterns have been observed both in evolved stars and in fainter turnoff and
subgiant cluster members.

It was soon recognized that the O-Na anticorrelation occurs thanks to the following
coincidence : at a similar temperature ($\sim 2.5 \times 10^7$~K), proton-captures
on $^{16}$O and $^{22}$Ne lead to the destruction of O and to the production of $^{23}$Na 
(\cite{DenissenkovDa1990}).
On the other hand the Mg-Al anticorrelation results from a sequence of proton-captures
followed by $\beta$-decays that transforms $^{24}$Mg into $^{25}$Mg, $^{26}$Mg
and finally $^{27}$Al 
(\cite{Langeretal1993}, \cite{LangerHoffman1995}) ;
this chain is only effective at temperatures higher than $\sim 7 \times 10^7$~K
due to the larger Coulomb barrier of Mg.
However the internal temperature of the scarcely evolved GC stars which exhibit 
the abundance anomalies is too low for these abundance variations to be intrinsic. 
This sustains the so-called {\sl self-enrichment} scenario according to which 
the abundance differences pre-existed in the material out of which the presently 
surviving stars formed. This requires the pollution of the intracluster gas
by a first generation of more massive and faster evolving stars 
\cite{CottrellDaCosta1981}\footnote{The fact that we see the same patterns in both
scarcely and strongly evolved stars, which have respectively very thin
and extremely deep convective envelopes, reveals primordial variations
instead of pollution on already formed stars.}.
These features have been observed in all the GCs where they have been looked for and 
appear thus to be intrinsic properties related to the cluster formation process itself.

Although the nuclear mechanisms that build up the anticorrelations are clearly described, 
the identification of the astrophysical site were they took place still remains a challenge.

\section{Polluting agents? AGB stars or massive rotating stars?}

It is claimed usually that massive AGB stars are responsible for the observed composition anomalies. 
However several custom-made detailed AGB models pointed out very severe drawbacks of the AGB pollution scenario.
These difficulties stem from the subbtle competition between hot bottom burning and third dredge-up.
This latter process does indeed contaminate the AGB envelope with the products of helium burning
and creates abundance patterns in conflict with the ones observed (\cite{Fenneretal04}, \cite{Charbonnel05}).

As an alternative \cite{PrantzosCharbonnel06} proposed the so-called 
{\sl {Winds of Massive Stars scenario}}
and suggested that hydrogen-burning in massive stars (i.e., with initial masses higher 
than $\sim$ 10 ~M$_{\odot}$) is at the origin of the abundance patterns\footnote{The idea that 
massive stars may be at the origin of some anomalies in the composition of GCs has been discussed 
by \cite{Norris04} and \cite{MM06} in order to explain the blue main sequence of the GC Omega Cen: 
the high helium content of the stars of that sequence could originate from the winds of massive stars, 
producing a large helium/metal ratio.}. In this case, the material ejected
in the interstellar matter by gently blowing winds of rapidly rotating massive stars does contribute 
to the formation of new low-mass stars. This idea is developped in great details and quantified 
by \cite{Decressinetal06} who also discusses qualitatively other advantages of that idea.

\section{Constraints on the IMF}
In the framework of the self-enrichment scenario for GCs, we present a new method to constrain 
the initial mass function (IMF) of the polluters (we refer the reader to \cite{PrantzosCharbonnel06}
for more details). 
We use the observed O/Na abundance distribution in NGC~2808 (\cite{Carretaetal06}) to derive 
the amount of polluted material with respect to that of original composition. 
We find that $\sim 30 \%$ of this GC stars have a pristine composition, while the remaining $70 \%$ 
has been contaminated to various degrees by H-burning products.

In view of the many uncertainties that enter this complex problem, 
we explore in some details two different types of self-enrichment scenarii differing in the 
composition of the polluter ejecta : 
Scenario I involves two clearly distinct stellar generation, the second of which is made 
exclusively from the nuclearly processed ejecta of the first one; in this case
the  ejecta of the polluter stars is processed at various degrees through H-burning.
Scenario II involves only one stellar generation, the low-mass stars of which are contaminated
on the making and to various degrees by extremely processed ejecta of their more massive
and rapidly evolving sisters. 
Also, we explore both current possibilities for the polluters, namely AGB stars (4-9~M$_{\odot}$) 
and massive stars (10-100~M$_{\odot}$). In each case we take the mass of H-processed ejecta 
as large as possible, in order to constrain the polluter IMF on one side : For AGB stars, 
we assume that all the mass outside the white dwarf remnant is processed exclusively 
through H-burning. For massive stars, we assume that all the mass outside the He-core 
has the required composition. 

We adopt a composite IMF, with an observationally derived part in the mass range 0.1-0.8~M$_{\odot}$
from \cite{ParesceDeMarchi00} and a power-law for higher masses with a slope X that we aim at 
constraining. Scenario I and Scenario II require respectively slopes X$<$ 0.8 and X$<$ 1.25 
if massive stars are the polluting agents, and X$<$ 0.15 and X$<$ 0.95 if AGB stars are the polluters. 
IMFs with the ``classical" Salpeter slope X=1.35 fail to satisfy the observational requirements in any case. 
 
The difficulty of the exercice stems on the fact that the parameter space is quite large. 
All our present assumptions are made in order to minimize the constraint on the IMF of
the polluter stars since their ejecta are used in the most efficient way by forming exclusively 
stars still alive today. If stars with initial masses higher than 0.8~M$_{\odot}$ were assumed 
to be also formed from the polluter ejecta, the corresponding mass required would be still larger 
and the IMF of the polluting agents even flatter than the ones we derived.

\section{Consequences for the amount of stellar residues}
Our study has also implications for the amount of dark objects (e.g., residues of stars 
with initial masses higher than 0.8~M$_{\odot}$) in GCs. 
The mass ratio of stellar residues to long-lived stars depends strongly 
on the assumption made about the mass range of the polluters, especially for flat IMFs as those
required to explain the abundance distribution in GCs. 

We find that the present number ratio of white dwarfs to long-lived stars, N$_{WD}$/N$_{MS}$,
should be around 0.2 if the polluters were AGB stars, and much smaller if the polluters 
were massive stars. These values are lower than the N$_{WD}$/N$_{MS}$ ratio infered by 
\cite{Richeretal02} in the case of the GC M4, and which is of the order of 1. 

The low number ratio of white dwarfs over low-mass stars we obtain does not necessary point to a fatal flaw 
for the self-enrichment scenarii. It may well be that the ejecta mass and the resulting number 
of second generation stars is smaller than assumed (this is in fact certainly the case in reality), 
in which case a N$_{WD}$/N$_{MS}$ ratio closer to the observationally infered one would be obtained.

\printindex

\begin{thebibliography}{99.}
\bibitem{Grattonetal2004} R.Gratton, C.Sneden, E.Carretta:
 ARAA \textbf{42}, 385 (2004)

\bibitem{Sneden2005} C.Sneden, IAU 228 228 on \textit{From Li to U : Element tracers of early
cosmic evolution}, Cambridge Univ.Press, Eds. Hill, Fran\c cois, Primas, p.337 (2005)

\bibitem{RamirezCohen2002} S.V.Ramirez, J.G.Cohen, J.G.:AJ \textbf{123}, 3277 (2002)

\bibitem{Ivansetal1999} I.I.Ivans, C.Sneden, R.P.Kraft, N.B.Suntzeff, V.V.Smith,
      G.E.Langer, J.P.Fullbright: AJ \textbf{118}, 1273 (1999)

\bibitem{DenissenkovDa1990} P.A.Denissenkov, S.N.Denissenkova S.N.: SvA Lett. \textbf{16}, 275 (1990)

\bibitem{Langeretal1993} G.E.Langer, R.Hoffman, C.Sneden, C.: PASP \textbf{105}, 301 (1993)

\bibitem{LangerHoffman1995} G.E.Langer, R.Hoffman, R.: PASP \textbf{107}, 1177 (1993)

\bibitem{CottrellDaCosta1981} P.L.Cottrell, G.S.Da Costa: ApJ \textbf{245}, L79 (1981)

\bibitem{Fenneretal04} Y.Fenner, S.Campbell, A.I.Karakas, J.C.Lattanzio, B.K.Gibson : MNRAS \textbf{353}, 789 (2004)

\bibitem{Charbonnel05} C.Charbonnel : IAU Symposium 228 on \textit{From Li to U : Element tracers of early
cosmic evolution}, Cambridge Univ.Press, Eds. Hill, Fran\c cois, Primas, p.347 (2005)

\bibitem{PrantzosCharbonnel06} N.Prantzos, C.Charbonnel : submitted to A\&A (2006)

\bibitem{Carretaetal06} E.Carretta, A.Bragaglia, R.G.Gratton, F.Leone, A.Recio-Blanco, S.Lucatello :
A\&A \textbf{450}, 523 (2006)

\bibitem{Decressinetal06} T.Decressin, G.Meynet, C.Charbonnel, N.Prantzos, S.Eckstr\"om :
submitted to A\&A (2006)

\bibitem{ParesceDeMarchi00} F.Paresce, G.De Marchi : ApJ \textbf{534}, 870 (2000)

\bibitem{Richeretal02} H.Richer, J.Brewer, G.Fahlman, et al : ApJ \textbf{574}, L151 (2002)

\bibitem{Norris04} J.E.Norris : ApJ \textbf{612}, L25 (2004)

\bibitem{MM06} A.Maeder, G.Meynet : A\&A \textbf{448}, L37 (2006)

\end{thebibliography}
\end{document}